\documentclass[aps,prx,reprint,twocolumn,superscriptaddress,floatfix,longbibliography]{revtex4-1}
\usepackage[T1]{fontenc}
\usepackage[latin9]{inputenc}
\usepackage{amsmath}
\usepackage{amssymb}
\usepackage{graphicx}
\usepackage{esint}
\usepackage[hyperindex,breaklinks]{hyperref}

\usepackage{color}
\usepackage{bm}
\usepackage{enumerate}
\usepackage{array}


\begin{document}

\title{Topological polaritons}

\author{Torsten Karzig}

\affiliation{Institute for Quantum Information and Matter, Caltech, Pasadena,
California 91125, USA}

\author{Charles-Edouard Bardyn}

\affiliation{Institute for Quantum Information and Matter, Caltech, Pasadena,
California 91125, USA}

\author{Netanel H. Lindner}

\affiliation{Physics Department, Technion, 320003 Haifa, Israel }

\affiliation{Institute for Quantum Information and Matter, Caltech, Pasadena,
California 91125, USA}

\author{Gil Refael}

\affiliation{Institute for Quantum Information and Matter, Caltech, Pasadena,
California 91125, USA}

\begin{abstract}
The interaction between light and matter can give rise to novel topological states. This principle was recently exemplified in Floquet topological insulators, where \emph{classical} light was used to induce a topological electronic band structure. Here, in contrast, we show that mixing \emph{single} photons with excitons can result in new topological polaritonic states --- or ``topolaritons''. Taken separately, the underlying photons and excitons are topologically trivial. Combined appropriately, however, they give rise to non-trivial polaritonic bands with chiral edge modes allowing for unidirectional polariton propagation. The main ingredient in our construction is an exciton-photon coupling with a phase that winds in momentum space. We demonstrate how this winding emerges from the finite-momentum mixing between $s$-type and $p$-type bands in the electronic system and an applied Zeeman field. We discuss the requirements for obtaining a sizable topological gap in the polariton spectrum, and propose practical ways to realize topolaritons in semiconductor quantum wells and monolayer transition metal dichalcogenides.
\end{abstract}
\maketitle

\section{Introduction}

The idea of creating topological photonic states was established in
2008 \cite{haldane_possible_2008,raghu_analogs_2008}. In a seminal
work, Haldane and Raghu proposed to generate the analog of quantum-Hall
states in photonic crystals with broken time-reversal symmetry. Shortly
thereafter, this concept was experimentally demonstrated for electromagnetic
waves in the microwave domain \cite{wang_observation_2009,poo_experimental_2011}.
Topological photonic states at optical frequencies, however, suffer
from the fact that the magnetic permeability is essentially $\mu=1$
in this regime \cite{landau_electrodynamics_1960}, which renders
the magneto-optical response allowing to break time-reversal symmetry
very weak. Proposed alternatives include the use of time-periodic
systems \cite{fang_realizing_2012,rechtsman_photonic_2013}, coupled
optical resonators or cavities \cite{cho_fractional_2008,koch_time-reversal-symmetry_2010,umucalilar_artificial_2011,hafezi_robust_2011,hafezi_imaging_2013,liang_optical_2013,jia_time_2013}, as well as metamaterials \cite{yannopapas_topological_2012,khanikaev_photonic_2013,chen_experimental_2014}.
Despite the accompanying experimental progress \cite{wang_observation_2009,poo_experimental_2011,rechtsman_photonic_2013,hafezi_imaging_2013}, the realization of photonic states with true topological protection at optical frequencies remains challenging. 

Here we pursue a new direction motivated by the concept of Floquet
topological insulators \cite{lindner_floquet_2011}, which rely on
the specific mixing of two topologically trivial fermionic bands made
possible by the absorption/emission of photons. In contrast to using
photons to generate a non-trivial topology, we ask whether one can
reverse this concept and create a non-trivial photon topology with
the help of electronic degrees of freedom. We show that this is indeed
possible by coupling photons to semiconductor excitons. Although both
of these ingredients are ordinary (non-topological) by themselves,
their combination can lead to new (quantum-Hall-like) topological
states of polaritons which we call, in short, ``topolaritons''.
We consider a setup where band mixing between $s$- and $p$-bands couples photons
and excitons in a non-trivial way. In combination with time-reversal
symmetry breaking for the underlying semiconductor (e.g., by a Zeeman
field), we then demonstrate that it is possible to generate a non-trivial topology.

\begin{figure}
  \begin{centering}
    \includegraphics[scale=0.23]{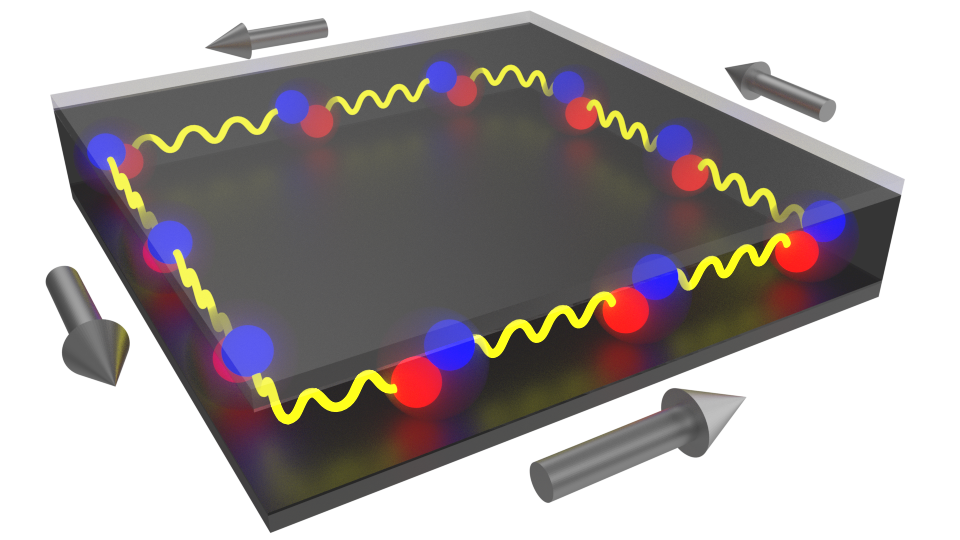}
  \end{centering}
  \caption{\label{fig:Schematic} Schematic view of a system with topological polaritons. A chiral polaritonic mode consisting of a mixture of bound particle-hole pairs (excitons) and photons is found at the edge of the system.}
\end{figure}

In contrast to quantum Hall states of fermions, the topology of the
bosonic (polariton) system that we consider is not a ground-state
property. Instead, it is characterized by topological bands in the
excitation spectrum. The main signature of this non-trivial topology
is a bulk gap in the excitation spectrum with chiral edge modes as
the only in-gap states (see Fig.~\ref{fig:Schematic}). These edge
modes provide a new realization of a controllable one-way waveguide
for photons \cite{figotin_electromagnetic_2003,yu_one-way_2008,wang_reflection-free_2008,takeda_compact_2008,wang_observation_2009,hadad_magnetized_2010,poo_experimental_2011}.
More conceptually, our proposal allows to realize topological photons
at optical frequencies and, to the best of our knowledge, constitutes
the first example of a topological hybrid state treating light
and matter degrees of freedom on the same footing. This is particularly interesting because finite interactions in the excitonic component open the perspective of interacting topological states of polaritons.

\section{Topolaritons}

Polaritons are superpositions of photons and excitons which can be
described by a Hamiltonian of the form
\begin{eqnarray}
  \hat{H} & = & \sum_{\mathbf{q}}\left[\omega_{q}^{{\rm C}}\hat{a}_{\mathbf{q}}^{\dagger}\hat{a}_{\mathbf{q}}
  +\omega_{q}^{{\rm X}}\hat{b}_{\mathbf{q}}^{\dagger}\hat{b}_{\mathbf{q}}
  +\left(g_{\mathbf{q}}\hat{b}_{\mathbf{q}}^{\dagger}\hat{a}_{\mathbf{q}}
  +{\rm H.c.}\right)\right]\,,\ \ \ \ \label{eq:exciton_photon_H}
\end{eqnarray}
where the operators $\hat{a}_{\mathbf{q}}^{\dagger}$ and $\hat{b}_{\mathbf{q}}^{\dagger}$
create photons and excitons with momentum $\mathbf{q}$, respectively.
We assume that the excitons and photons are both confined to two dimensions,
i.e., $\mathbf{q}=(q_{x},q_{y})$. For the excitons, this can be achieved using a quantum well, while photons can be trapped using a microcavity
or waveguide. The exciton dispersion $\omega_{q}^{{\rm X}}=q^{2}/2m_{{\rm X}}+\omega_{0}^{{\rm X}}$
(setting $\hbar=1$) describes its center-of-mass motion, while the
energy gap $\omega_{0}^{{\rm X}}$ for creating an exciton is given
by the difference between the bare particle-hole excitation gap and
the exciton binding energy. We denote the dispersion of the cavity
photon by $\omega_{q}^{{\rm C}}$.

The crucial ingredient for generating topolaritons is the exciton-photon
coupling $g_{\mathbf{q}}$ which describes the creation of an exciton
by photon absorption and vice versa. Here we require $g_{\mathbf{q}}$ to wind according to 
\begin{equation}
  g_{\mathbf{q}}=g_{q}{\rm e}^{{\rm i}m\theta_{\mathbf{q}}},\label{eq:winding}
\end{equation}
where $g_{q}$ is the amplitude of the exciton-photon interaction
(or Rabi frequency), $m$ is a non-zero integer, and $\theta_{\mathbf{q}}$
denotes the polar angle of $\mathbf{q}$.

\begin{figure}
  \begin{centering}
  \includegraphics[scale=0.95, bb=0bp 0bp 252bp 133bp]{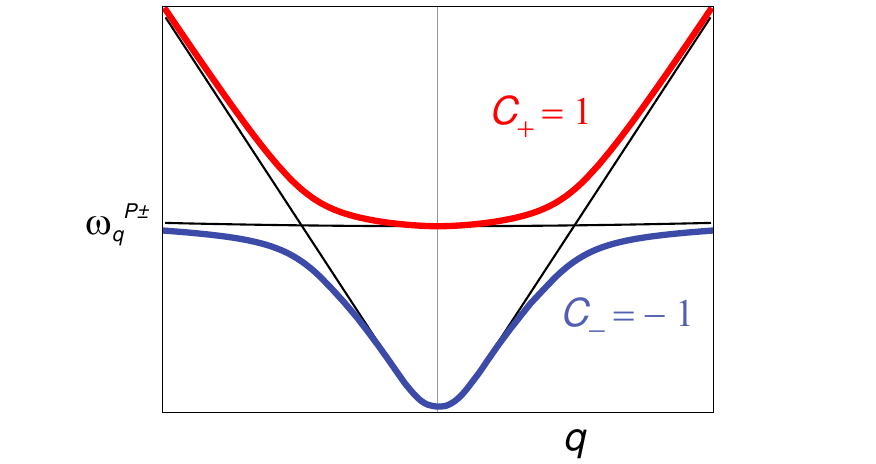}
  \end{centering}
  \caption{\label{fig:Polariton-spectrum} Schematic polariton spectrum $\omega_{q}^{{\rm P}\pm}$. The upper and lower polariton bands are drawn in red and blue, while the uncoupled bands ($g_{q}=0$) are shown as thin black lines for comparison. Due to the winding structure of the coupling, the two polariton bands acquire a non-trivial topology with finite Chern numbers $C_\pm=\pm1$ in the case of $m=1$ {[}see Eq.~\eqref{eq:winding}{]}.The associated edge modes are indicated by a dashed purple line.}
\end{figure}

To reveal the non-trivial topology, we diagonalize the Hamiltonian
\eqref{eq:exciton_photon_H} in terms of polariton operators 
\begin{equation}
\hat{P}_{\mathbf{q}}^{\pm}=\mathbf{e}_{\mathbf{q}}^{P\pm}\cdot\left(\begin{array}{c}
\hat{a}_{\mathbf{q}}\\
\hat{b}_{\mathbf{q}}
\end{array}\right)
\end{equation}
with spectrum $\omega_{q}^{{\rm P}\pm}=\frac{1}{2}\left(\omega_{q}^{{\rm C}}+\omega_{q}^{{\rm X}}\right)\pm\frac{1}{2}\big[\left(\omega_{q}^{{\rm C}}-\omega_{q}^{{\rm X}}\right)^{2}+4g_{q}^{2}\big]^{1/2}$, where the $\pm$ sign refers
to the upper and lower polariton band, respectively (see Fig. \ref{fig:Polariton-spectrum}). The vectors $\mathbf{e}_{\mathbf{q}}^{P\pm}$
describe the relative strength between the photon and exciton components
of the polariton wavefunction and can be interpreted as a spinor. The non-trivial form of the coupling \eqref{eq:winding} leads to
a winding of this spinor of the form 
\begin{equation}
  \mathbf{e}_{\mathbf{q}}^{P\pm}=\frac{1}{\sqrt{2}}\left(\begin{array}{c}
  \pm{\rm e}^{-{\rm i}m\theta_{\mathbf{q}}}\sqrt{1\pm\beta_{q}}\\
  \sqrt{1\mp\beta_{q}}\end{array}\right)\,,
  \label{eq:eigenvectors}
\end{equation}
where $\beta_{q}=\left(\omega_{q}^{{\rm C}}-\omega_{q}^{{\rm X}}\right)/\big[\left(\omega_{q}^{{\rm C}}-\omega_{q}^{{\rm X}}\right)^{2}+4g_{q}^{2}\big]^{1/2}$.
The spinor of the lower polariton band ``flips'' from photonic to
excitonic (vice versa for the upper polariton) far away from the resonance
as described by the limits $\beta_{q=0}=-1$ and $\beta_{q\rightarrow\infty}=1$.
Combined with the winding ${\rm e}^{{\rm i}m\theta_{\mathbf{q}}}$,
this flip leads to a full wrapping of the unit sphere by the spinor
$\mathbf{e}_{\mathbf{q}}^{P\pm}$, thereby leading to a non-trivial
topology in full analogy to fermionic topological systems. This can
be confirmed by calculating the Chern number of the upper and lower
polariton bands. Indeed, we find $C_{\pm}=\pm m$. Note that the mechanism leading to non-trivial topology is distinctly different from the original proposal by Haldane and Raghu \cite{haldane_possible_2008,raghu_analogs_2008}. Here  the non-trival Berry curvature arises directly from the (winding) hybridization between two ordinary exciton and photon bands, not from the gapping out of symmetry-protected band touchings (e.g., Dirac cones) in the photonic spectrum \footnote{In fact, the resulting edge modes in our case are predominantly of excitonic character, which stresses that the semiconductor degrees of freedom are more important for the non-trivial topology then simply providing an effective magneto-optical coupling}.

A consequence of the non-trivial Chern number is the presence of chiral
polaritonic edge modes. In our setting, edges are defined by the confinement
of the excitons and photons. While excitons are naturally restricted
to the quantum well, photons can be confined, e.g., by using reflecting
mirrors or a suitable photonic bandgap at the edges of the system.
Another possibility would be to confine the photons in a dielectric
slab waveguide ending at the system edges.

Besides the existence of bands with non-trivial Chern numbers, the
stability of the edge modes also requires a global energy gap (i.e.,
present for all momenta) between the upper and lower polariton bands.

\begin{figure}
  \begin{centering}
  \includegraphics[bb=0bp 0bp 250bp 155bp]{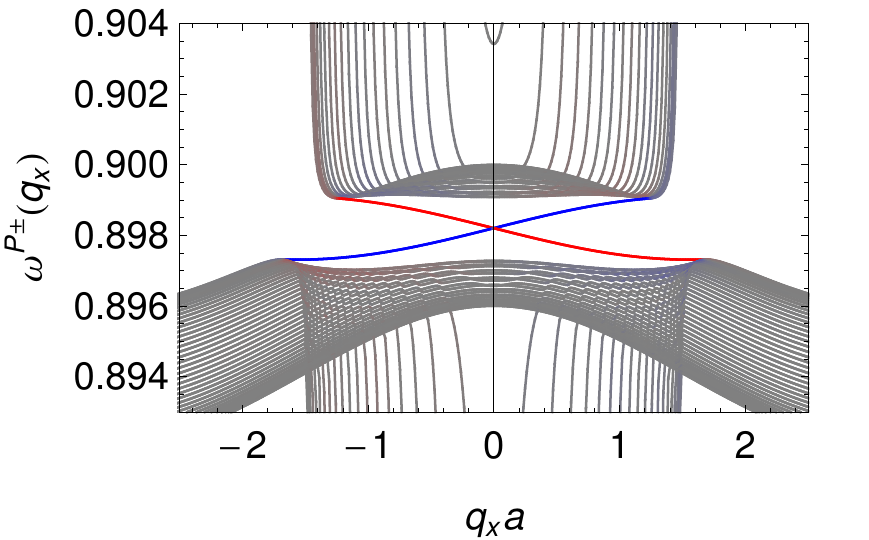}
  \end{centering}
  \caption{\label{fig:Polariton-edge-modes} Topological polariton spectrum. Spectrum obtained for a tight-binding lattice analog of the Hamiltonian~\eqref{eq:exciton_photon_H} on a square lattice with periodic (open) boundary conditions in the x-direction (y-direction). Two edge modes are found within the gap, each localized at one of the system boundaries. The color scale encodes the edge localization, where red (blue) indicates a large weight of the wavefunction at the edge at $y=0$ ($y=L$). For the construction of the lattice model, we choose quadratic dispersions $q^{2}/2m_{X}+\omega_{0}^{X}$ and $q^{2}/2m_{C}$ for the exciton and photon, respectively, replacing $q_{x,y}^{2}\rightarrow2-2\cos(q_{x,y})$ (note that a quadratic photon dispersion with $m_{C}=q_{{\rm res}}/2c$ correctly reproduces the original linear dispersion close to resonance at $q_{{\rm res}}$). Similarly, we model the winding coupling by $g_{\mathbf{q}}=g[\sin(q_{x})+\mathrm{i}\sin(q_{y})]$. The system size is $40\times40$ (with lattice constant $a=1$) and the other relevant parameters are given by $m_{{\rm X}}=-500,$ $m_{{\rm C}}=1$, $\omega_{0}^{{\rm X}}=0.25$, and $g=0.01$.}
  \label{fig:numerical_negative_mass}
\end{figure}

Above we assumed that the bare exciton and photon dispersions are far (negatively) detuned at $q=0$. In that case the energy of the lower polariton branch essentially coincides with the bare exciton dispersion for low momenta. Since the lower polariton branch approaches the bare exciton dispersion
for large $q$, it thus appears impossible to open a gap for a positive exciton mass. Although a small gap can in principle be opened by considering a negative exciton mass, as depicted in Fig.~\ref{fig:Polariton-edge-modes}, we present below a more realistic and efficient way to realize topolaritons. We first discuss how to obtain a winding exciton-photon coupling, and show that a sizable topological gap can be opened in the presence of a periodic exciton potential.

\section{Realizing topolaritons}

\begin{figure}
	\begin{centering}
		\includegraphics[scale=0.9]{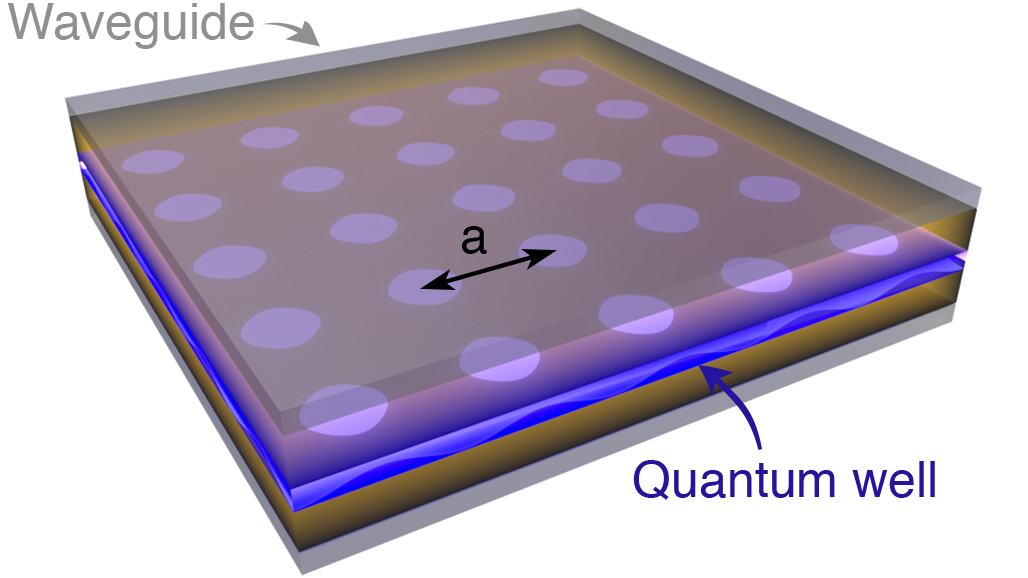}
	\end{centering}
	\caption{\label{fig:setup}
		Schematic experimental setup for realizing topolaritons. A quantum well (in blue) is embedded in a photonic waveguide (or cavity). The boundary of the light blue disks represent equipotential lines of the periodic exciton potential with lattice constant $a$ (see Sec.~\ref{sec:top_gap}).}
\end{figure}

\subsection{Winding coupling}
Promising candidates for the realization of topolaritons are semiconductor quantum wells embedded in photonic waveguides or microcavities (Fig.~\ref{fig:setup}).
One of the main requirements is the presence of a single (bright) two-dimensional excitonic mode. Here we consider the standard case of excitons formed from $s$-type electronic states in the conduction band and $p$-type heavy-hole states in the valence band. The essential part of the quantum well $k\!\cdot\!p$ Hamiltonian describing these particle and hole bands can be expressed in the form~\cite{winkler_spin-orbit_2003}
\begin{align}
  H_{{\rm QW}} & =\left(M+\frac{k^{2}}{2m}\right)\sigma_{z}
  +A\left(k_{x}\sigma_{x}+k_{y}\sigma_{y}\right)\,,\label{eq:quantum well}
\end{align}
where $M>0$ determines the size of the (trivial) bandgap, $\sigma_{i}$ are Pauli matrices corresponding to states with different total angular momentum ($J_z=-1/2$ and $J_z=-3/2$),
and $A$ describes the finite-momentum interband mixing.

The original $k\!\cdot\!p$ Hamiltonian also includes the time-reversed version of $H_{{\rm QW}}$ (involving $J_z=1/2$ and $J_z=3/2$ states). Here we break time-reversal symmetry and isolate a single of these blocks by applying an external magnetic field in the $z$-direction. A finite Zeeman field can shift the exciton energy when the conduction and valence bands have different g-factors. Below we assume that the resulting energy splitting between excitons belonging to different blocks is large enough so that one can focus on a single one of them. We note that this strict separation is only justified provided that the Zeeman splitting is larger than the exciton-photon coupling. Our results, however, remain valid for much weaker Zeeman fields (only exceeding the size of the topological gap). Taking into account both excitons, we find two topological gaps energetically separated by the Zeeman splitting, with respective edge states of opposite chirality (see Supplemental Material). We remark that orbital effects of the magnetic field can be neglected as long as the magnetic length is larger than the size of the excitons, which we shall assume in what follows.

When adding Coulomb interactions to the quantum-well Hamiltonian
\eqref{eq:quantum well}, excitons form as bound states of conduction-band
particles and valence-band holes. The Bohr radius $a_{0}$ of such
``particle-hole atoms'' is typically of the order of $1-10$nm, with
binding energies of the order of $10-100$meV (see, e.g., Ref.~\cite{deng_exciton-polariton_2010}).
Expressed in terms of the creation operators $\hat{c}_{c,{\mathbf k}}^\dagger$ and  $\hat{c}_{v,{\mathbf k}}$ for the conduction- and valence-band state, the corresponding exciton operator takes the form
\begin{equation}
  \hat{b}_{\mathbf{q}}^{(n)\dagger}
  =\sum_{\mathbf{k}}\phi_{n}\left(\mathbf{k}\right)
  \hat{c}_{c,\mathbf{k}+\mathbf{q}/2}^{\dagger}\hat{c}_{v,\mathbf{k}-\mathbf{q}/2}
  \label{eq:exciton-1}
\end{equation}
and obeys bosonic commutation relations for low exciton densities
(note that we assume equal masses for particles and holes, for simplicity).
The exciton wavefunctions $\phi_{n}(\mathbf{k})$ are (the Fourier
transform of) hydrogen-atom-like wavefunctions in relative coordinates for the bound particle-hole pairs.

For the photonic part of the Hamiltonian, we assume that the photons are confined to two dimensions by a waveguide or a microcavity (the semiconductor quantum well being either in direct proximity or
inside the latter). The two-dimensional photonic modes are either of transverse-electric (TE) or transverse-magnetic (TM) nature. To open a gap in the polariton spectrum, we want to target a regime where a single photonic mode couples to a single excitonic mode. This can be realized by using a photonic crystal with a suitably tailored TM bandgap (see e.g., Ref.~\cite{joannopoulos_photonic_2011}) where the presence of TM modes within the topological gap is avoided.

We remark that it is in principle also possible to observe topological features in the absence of a TM gap, in the limit of large in-plane momentum around the exciton-photon resonance $q_{\rm res}$. In this regime the electric field of the TM mode points predominantly out of the plane. Since two-dimensional excitons only couple to in-plane electric fields, the coupling to the TM modes vanishes in this limit, suppressed by a factor $1/(q_{\rm res}d)$, where $d$ is the thickness of the waveguide or microcavity. In practice, the (finite but small) coupling of excitons to TM modes will lead to some mixing
of the resulting chiral edge states with the TM modes, which spoils true topological protection. For theoretical clarity, we focus on the TE mode in the following, and discuss the details of the TE/TM interplay in the Supplemental Material.

The quantized vector potential describing the TE modes takes the form
\begin{equation}
  e\hat{\mathbf{A}}(\mathbf{r},t)
  =\int\frac{{\rm d}q^{2}}{\left(2\pi\right)^{2}}F_{\mathbf{q}}
  \left({\rm \mathbf{e}}_{\mathbf{q}\perp}\hat{a}_{\mathbf{q}}
  {\rm e}^{{\rm i}(\mathbf{q}\cdot\mathbf{r}-\omega_{\mathbf{q}}t)}
  +{\rm H.c.}\right),
\end{equation}
where $\hat{a}_{\mathbf{q}}^{\dagger}$ creates a photon of momentum
$\mathbf{q}$, $\mathbf{e}_{\mathbf{q}\perp}=(-q_{y},q_{x})/q$ describes
the direction of the in-plane electric field perpendicular to the
propagation direction $\mathbf{q}$, and $F_{\mathbf{q}}=\left[e^{2}/\left(2\epsilon w\omega_{\mathbf{q}}^{{\rm C}}\right)\right]^{1/2}$,
where $e$ denotes the electron charge, $\epsilon$ the dielectric
constant, and $w$ the width of the waveguide/microcavity in the $z$-direction.
The coupling of the photons to the quantum well can be incorporated
by the standard substitution \textbf{$\mathbf{k}\rightarrow\mathbf{k}+e\hat{\mathbf{A}}$}
in the Hamiltonian~\eqref{eq:quantum well}. The mixing between the $s$- and $p$-bands in the quantum well, combined with the locking of the electric-field direction (${\mathbf e}_{{\mathbf q}\perp}$) to ${\bf q}$, then leads to a winding exciton-photon coupling. For s-wave excitons, and provided that $A/(Ma_0) < 1$ (see Appendices), we find an exciton-photon Hamiltonian of the form \eqref{eq:exciton_photon_H}, with a winding coupling
\begin{equation}
  g_{\mathbf{q}} \approx
  -{\rm i}\sqrt{\frac{2}{\pi}}\frac{A}{a_{0}}F_{\mathbf{q}}\mathrm{e}^{-\mathrm{i} m \theta_{\mathbf{q}}}\,,
  \label{eq:winding_g}
\end{equation}
where $m = 1$. Intuitively, this single winding can be understood form the fact that angular momentum conservation is required when mixing excitons with total angular momentum projection $J_z=1$ with the (linearly polarized) TE mode of $J_z=0$. Higher winding numbers $m = \pm2$ can in principle be achieved starting from excitons with $J_z=\pm1$ coupled to photons with helicity $\mp$1. We detail this alternative scheme in the Supplemental Material. As far as higher winding numbers are concerned, it would also be interesting to extend our proposal to systems in which excitons already have a non-trivial topological nature~\cite{yuen-zhou_topologically_2014,pikulin_interplay_2014,budich_time_2014}.
 
 Note that despite its winding form, the coupling of Eq.~(\ref{eq:winding_g}) originates from the same dipole moment as the standard (constant) coupling $\Omega$ of $J_z=+1(-1)$ excitons to right(left)-handed polarized light. This can be seen from the small momentum limit,  where the right-handed polarized mode is given by $a^+_\mathbf{q}=(a^{\rm TM}_\mathbf{q}-\mathrm{i}a^{\rm TE}_\mathbf{q})\mathrm{e}^{-\mathrm{i}\theta_\mathbf{q}}/\sqrt{2}$ (see Supplemental Material), yielding $|g_\mathbf{q}|=\Omega/\sqrt{2}$.

\subsection{Finite topological gap} \label{sec:top_gap}
\begin{figure}
	\begin{centering}
		\includegraphics[scale=0.9, bb=0bp 0bp 259bp 163bp]{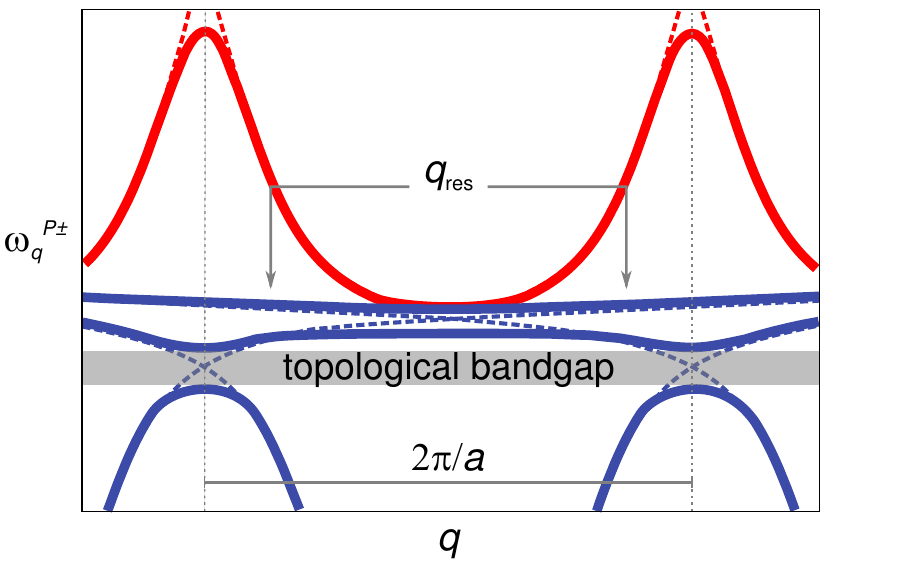}
	\end{centering}
	\caption{\label{fig:schematic periodic potential}
		Scheme for opening a topological gap using a periodic potential. The solid (dashed) blue and red lines correspond to the dispersion of the lower and upper polaritons with (without) a periodic potential in a simplified one-dimensional scenario. As long as $\pi/a>q_{{\rm res}}$, the lowest polariton band includes the winding around the resonance and the gap opened at the Brillouin-zone boundary is of topological nature.}
\end{figure}

The remaining ingredient for chiral polaritonic edge modes is a finite topological gap. One way to open such a gap is to introduce a periodic potential
in the coupled exciton-photon system (see Fig.~\ref{fig:setup}).
While periodic potentials for photons are naturally provided by photonic
crystals \cite{yablonovitch_inhibited_1987}, excitonic analogs can
be realized, e.g., by applying strain to the quantum well (see Ref.~\cite{carusotto_quantum_2013} and references therein), or using surface acoustic waves~\cite{de_lima_phonon-induced_2006,cerda-mendez_exciton-polariton_2013}.
The presence of a periodic potential with period $a$ introduces a
Brillouin zone for the polariton spectrum. The coupling of the backfolded
bands at the Brillouin-zone boundary then allows to open a gap (see
Fig.~\ref{fig:schematic periodic potential}).

To understand the topological nature of such a gap, we recall that
most of the winding responsible for the non-trivial topology takes
place around the circle of radius $q_{{\rm res}}$ corresponding to the exciton-photon resonance. For
$\pi/a> q_{{\rm res}}$, we thus expect the same non-trivial Chern
number as in the absence of a periodic potential, while for $\pi/a< q_{{\rm res}}$
the lowest polariton band should be topologically trivial. To achieve a large
topological gap, the Brillouin-zone boundary should correspond to
a momentum of the order of (but larger than) $q_{{\rm res}}$.

Note that opening a topological gap in the exciton-dominated polariton regime (i.e., at $q>q_{{\rm res}}$), requires a periodic exciton potential. In particular, it is not possible to use purely photonic periodic potentials. The latter can only gap out the excitonic part via the (winding) 
exciton-photon coupling $g_{\mathbf{q}}$. This results in an additional
winding coupling between the excitonic bands at the Brillouin-zone
boundary which cancels the original winding and makes the gap between
the lowest polariton bands topologically trivial. Note that due to the weaker effect of photonic potentials on the exciton-dominated regime, including periodic exciton and photon potentials will in general still lead to topological gaps \footnote{In particular, we find well-defined topological gaps in the regime of a periodic-potential-induced TM gap as discussed in the Supplemental Material.}. To simplify the discussion, here we focus on periodic potentials that are purely excitonic and discuss the case of both exciton and photon potentials in the Supplemental Material.

Figure~\ref{fig:schematic periodic potential} provides an intuitive
picture of our gap-opening scheme in a simplified one-dimensional
case. For the actual two-dimensional Brillouin zone, one essentially recovers this scenario for any cut passing through the center of the Brillouin zone (see Appendices). The optimal way to open a topological gap is to maximize the overlap between the gaps obtained along each possible cut. This occurs when the Brillouin-zone geometry is as circular as possible (e.g., hexagonal).

\subsection{Numerical results}
\begin{figure}
  \begin{centering}
  \includegraphics[bb=0bp 0bp 250bp 160bp]{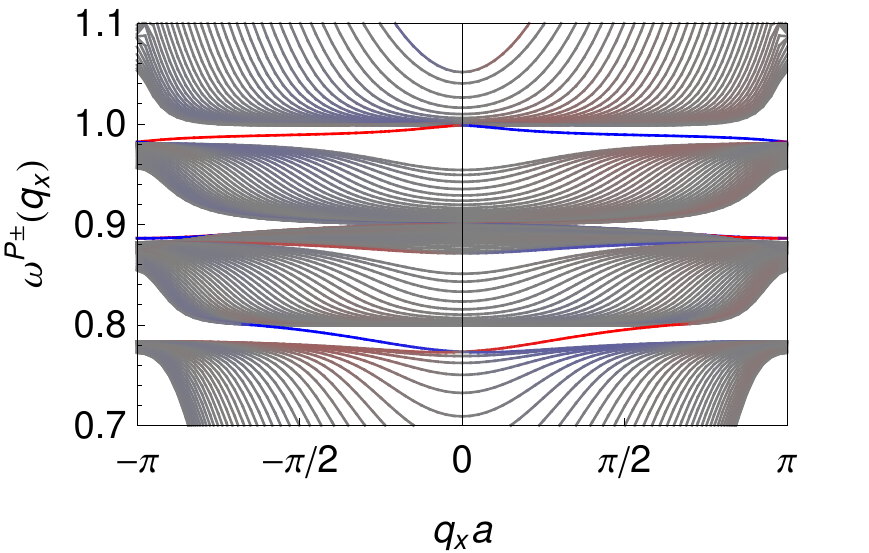}
  \end{centering}
  \caption{\label{fig:periodic_lattice_spectrum} Topological polariton spectrum with a periodic potential. The spectrum is obtained for a finite-size lattice-version of the Hamiltonian \eqref{eq:exciton_photon_H} with an additional periodic potential of period $a$ and exhibits in-gap chiral edge modes (colored). The color scale encodes the edge localization, where red (blue) indicates a large weight of the wavefunction at the edge at $y=0$ ($y=L$). The system size is $40\times40$ (with lattice constant $a=1$) and the other relevant parameters are given by $m_{{\rm X}}=1\cdot10^{5},$ $m_{{\rm C}}=1$, $\omega_{0}^{X}=0.9,$ $g=0.1$,$V_{{\rm X}}=0.05$.}
\end{figure}

To study the edge modes numerically, we start from the same lattice
model as in Fig.~\ref{fig:numerical_negative_mass} and set the lattice
constant to $a/2$. We then introduce a periodic potential of period
$a$ by alternating the on-site energy between neighboring sites of
the square lattice. Figure~\ref{fig:periodic_lattice_spectrum} shows
the resulting spectrum, which clearly demonstrates the presence of polaritonic
edge modes. Since the topological gap $\Delta$ is controlled by the
periodic potential, we find that the typical length scale for the edge-mode localization is of the order of $a$ (see Fig.~\ref{fig:periodic_lattice_spectrum}, where the extend of the edge modes in momentum space is of the order of $\pi/a$). The size of the gap then determines the edge-mode velocity $v\sim a\Delta/\pi$. Since the exciton velocity is essentially
negligible as compared to the photon velocity $c$, the edge-mode
velocity in turn dictates a ratio $v/c$ between the photon and exciton
components of the edge mode.

\subsection{Practical realization} What are the necessary features and parameters for a quantum well to host topolaritons? Most importantly, it must support free excitons. In addition, the exciton-photon coupling $g_{\mathbf{q}}$ must be as large as possible since it defines, along with the periodic exciton potential, the relevant gaps in the polariton spectrum. Equation~\eqref{eq:winding_g} ties $g_{\mathbf{q}}$ to the ratio of the finite-momentum interband mixing amplitude $A$ to the exciton Bohr radius $a_0$, and is valid provided that $A/a_0 < M$ (see also Eq.~\eqref{eq:quantum well}). Together, the above conditions thus require a large dipole moment (controlling $A$) and a bandgap larger than $A/a_0$. The size of the topological gap is ultimately limited by the strength of the periodic exciton potential and by the exciton Zeeman splitting. A crucial requirement for the resulting gap is that it exceeds the inverse polariton lifetime, such that the topological edge modes are well defined despite the decay-induced broadening of the polariton states.

We demonstrate the feasibility of our proposal using CdTe-based quantum wells as a guideline. For an exciton energy $\omega_0^{\rm X} = 1.6 {\rm eV}$ and a dielectric constant $\epsilon = 8$, the condition $\pi/a \sim q_{\rm res}$ is fulfilled for a periodic exciton potential with lattice constant $a \approx 150 {\rm nm}$. CdTe-based quantum wells are a standard platform to reach the strong-coupling limit. Recently, exciton-photon couplings of up to $\Omega=|g_\mathbf{q}|\sqrt{2}=55 \mathrm{meV}$ have been reached using multiple CdTe-based quantum wells ($\Omega=5.4$meV for a single quantum well) embedded in a photonic-crystal-based microcavity~\cite{jiang_photonic_2014}. For such large exciton-photon couplings, the size of the topological gap would essentially only be limited by the strength of the periodic exciton potential and the exciton Zeeman splitting. For more moderate exciton-photon couplings $\Omega=20\mathrm{meV}$ and an optimal triangular-lattice potential of depth $1{\rm meV}$, a topological gap of $0.3{\rm meV}$ can be obtained. Such a gap would be large enough to achieve stable polariton edge modes, since typical polariton lifetimes are of the order of tens of picoseconds ($1/10{\rm ps} = 0.07{\rm meV}$)~\cite{deng_exciton-polariton_2010}. With a typical exciton g-factor of about 2 \cite{sirenko_electron_1997}, the Zeeman energy would not be a limiting factor for the above topological gap of $0.3$meV when applying magnetic fields of $2.5 \mathrm{T}$ or larger.

\begin{figure*}
\begin{centering}
  \includegraphics[bb=0bp 0bp 520bp 198bp]{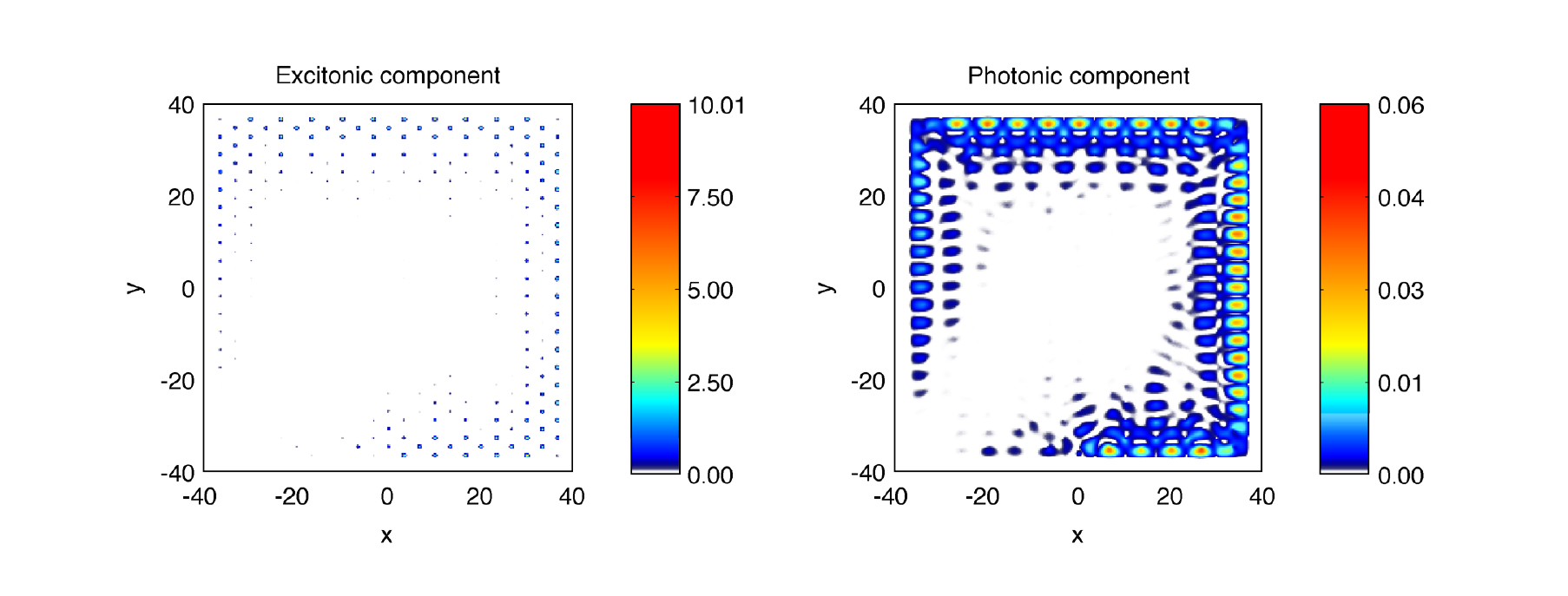}
  \end{centering}
  \caption{\label{fig:GPE}
   Topolariton edge modes. The figure depicts the typical intensities of the exciton and photon fields obtained when pumping a finite square system optically at the center of its lower edge, shown here a time $t=4\cdot10^{3}$ after switching on the laser for an triangular-lattice exciton potential of strength $V_{{\rm X}}=-0.05$ and lattice constant $a=3.86$. Units are defined here by setting $\omega_{0}^{{\rm X}}$ and the speed of light to unity (other relevant parameters are $m_{{\rm X}}=1\cdot10^{3},\ g_{q}=0.1$, the pumping frequency $\omega_{{\rm p}}=0.856$, and a polariton decay rate $\gamma=5\cdot10^{-4}$). A counter-clockwise chiral edge mode is clearly visible, with an exciton component strongly peaked at the minima of the periodic potential. The apparent difference between the distance traveled by the exciton and the photon fields is due to the color scale which makes it easier to resolve the more extended photon fields. Exciton and photon fields both travel together as a unique polaritonic chiral edge mode.}
\end{figure*}

A promising alternative for realizing topolaritons is provided by monolayers of transition metal dichalcogenides (TMDs). These atomically thin two-dimensional materials with graphene-like (honeycomb) lattice structure have attracted a lot of interest in recent years owing to their coupled spin and valley degrees of freedom, as well as their large direct bandgap ($1$-$2{\rm eV}$) and interband mixing (see, e.g., Ref.~\cite{xiao_coupled_2012}). In the presence of an external magnetic field which lifts their valley degeneracy~\cite{macneill_breaking_2015,srivastava_valley_2015,aivazian_magnetic_2015}, their electronic properties can be described by an effective two-band Hamiltonian of the form~\eqref{eq:quantum well} with gap and interband-mixing parameters $M$ and $A$ as large as $0.5$-$1{\rm eV}$ and $3.5$-$4.5{\rm eV}\,$\AA, respectively~\cite{xiao_coupled_2012,cai_magnetic_2013,rose_spin-_2013}. Exciton-polaritons in the strong-coupling regime have also been achieved, with exciton binding energies close to $1{\rm eV}$ and a Bohr radius of the order of $1{\rm nm}$~\cite{liu_strong_2014}. With such parameters, the size of the exciton-photon coupling for a monolayer TMD would be of the order of $20\mathrm{meV}$,  leading to similar topological gaps as for the CdTe multi-quantum-well setups discussed above ($\Delta=0.3$meV for periodic exciton potentials of depth 1meV). Such a gap would require a magnetic field of about $1.5$T to lift the valley degeneracy~\cite{macneill_breaking_2015,srivastava_valley_2015,aivazian_magnetic_2015}, corresponding to a regime where the magnetic length is much larger than the exciton Bohr radius, as desired. We remark that TMDs would also provide a promising platform to realize topolaritons with winding number $|m| = 2$ using circularly polarized light, as discussed in the Supplemental Material. In that case the circular polarization of light automatically selects one of the two valleys, thus obviating the need for a magnetic field~\cite{cao_valley-selective_2012,zeng_valley_2012,mak_control_2012}.

For the photonic part, as mentioned above, the most practical way towards topolaritons is a two-dimensional cavity with a tailored TM gap (e.g., by using a photonic crystal). This avoids the difficult large in-plane momentum regime and leads to a true topological protection of the edge mode.  Note that polaritons have been observed in photonic crystals \cite{bajoni_exciton_2009}, which were also recently used to reach very strong exciton-photon couplings \cite{jiang_photonic_2014}.

\section{Probing topolaritons}

The non-trivial topology of topolaritons manifests itself through
the existence of chiral edge modes in the excitation spectrum. Probing
these modes requires both a means to excite them, and the ability to 
observe them. Due to their mixed exciton-photon nature, chiral edge
polaritons can be excited either electronically or optically. Possible
electronic injection processes include exciton tunneling \cite{lawrence_exciton_1994}
or resonant electron tunneling \cite{cao_direct_1995}. More commonly,
polaritons can be created by addressing their photonic component using
a pump laser. Either way, the spectral resolution of the injection
process should ideally be smaller than the topological gap in order
to single out the chiral edge modes of interest from the rest of the
spectrum. A lower spectral resolution can be compensated by better
spatial focusing at the edge where the bulk modes have less weight
using, e.g., laser spot sizes $<a$.

Once excited, topological polaritonic edge modes propagate chirally, which results in clear differences when observing them upstream and
downstream along the edge. Although these chiral edge modes are protected
from backscattering, exciton and photon losses translate as a decay of their wavefunction over time. Such dissipation is both harmful
and useful: On one hand, the finite decay rate eventually spoils the
stability of the edge mode and should not exceed the size of the topological
gap for the edge modes to be well-defined. On the other hand, photon
losses provide a way to probe the existence of edge modes. For example,
photons confined in a waveguide cannot escape when propagating along
a smooth edge. Sharp turns (such as corners) would allow to couple
to the outside continuum of modes, which could be used to locally
detect or inject topological polaritonic edge modes.

To incorporate pumping and loss effects and to complement the lattice-model
results of Fig.~\ref{fig:periodic_lattice_spectrum}, we simulated what
we envision as a typical probe experiment using a driven-dissipative
Gross-Pitaevskii equation based on the continuum model of our proposal
(see Supplemental Material). We considered a scenario where a
continuous-wave pump laser with a frequency $\omega_{{\rm p}}$ set
within the topological gap is switched on at some initial time $t=0$
and focused onto a spot of diameter $\approx a$ at the edge. Figure~\ref{fig:GPE}
shows the time-evolved exciton and photon fields obtained in that
case. The photonic part of the edge modes can in principle be easily
retrieved by detecting the light leaking out of the system. For example,
the photon component shown in Fig.~\ref{fig:GPE} would be observed in a system with uniform losses, by detecting photons escaping from the cavity in the $z$-direction.
We remark that exciton-exciton interactions are expected to be present
in typical experimental realizations (see, e.g.,~Ref. \cite{deng_exciton-polariton_2010}).
Our numerical studies indicate that the edge modes remain stable in
the presence of weak interactions (i.e., for interactions leading
to a blueshift smaller than the topological gap), as expected from
their chiral nature (see Supplemental Material and Video).

\section{Conclusions}

In this manuscript, we have shown that topological polaritons (``topolaritons'')
can be created by combining ordinary photons and excitons. The key
ingredients of our scheme are a winding exciton-photon coupling
and a suitable exciton potential to open a topological gap. Promising
candidates for the realization of the winding coupling are two-dimensional
TE-polarized photonic modes coupled to semiconductor quantum wells or monolayer transition metal dichalcogenides under a finite magnetic field. Although combining all necessary
ingredients may be experimentally challenging, we emphasize that all
underlying requirements have already been achieved. Realizing topolaritons
thus seems within experimental reach. In light of the importance of
a Brillouin zone for the opening of the gap, it would be interesting
to study whether our proposal can be extended to systems known to
exhibit strong polariton potentials, such as lattices of coupled micropillars
\cite{jacqmin_direct_2014}.

Topolaritons manifest themselves through chiral edge modes which are, by definition, protected against backscattering. The directionality of these one-way channels can in principle easily be reversed, either by flipping the direction of the magnetic field or by addressing their time-reversed partners appearing at different energies. This makes for a versatile platform for the directed propagation of excitons and photons, with the possibility of converting between the two.

An important feature of topolaritons is their strong
excitonic component, which allows for interactions to come into play. Interaction effects are notoriously hard to achieve in purely photonic topological systems
\cite{carusotto_quantum_2013}. Here they could provide, in particular, an alternative and more flexible way to realize the periodic exciton potential required in our proposal. Indeed, one may envision to create an effective potential by injecting
a different exciton species with a periodic density profile \cite{amo_light_2010}. Interactions may also lead to novel
avenues towards the observation of other intriguing topological phenomena
such as non-equilibrium fractional quantum Hall effects \cite{cho_fractional_2008,umucalilar_fractional_2012,hafezi_non-equilibrium_2013}.

More broadly, our route to obtaining topological polaritons reveals a generic approach for achieving non-trivial topological states by mixing trivial bosonic components. We expect extensions of this idea to other physical systems to lead to the realization of yet more surprising bosonic topological phenomena.

\begin{acknowledgments}
We are grateful to Bernd Rosenow, Alexander Janot, and Andrei Faraon
for valuable discussions. This work was funded by the Institute for
Quantum Information and Matter, an NSF Physics Frontiers Center with
support of the Gordon and Betty Moore Foundation through Grant GBMF1250, NSF through DMR-1410435, the David and Lucile Packard Foundation, 
the Bi-National Science Foundation and I-Core: the Israeli Excellence
Center ``Circle of Light'', and Darpa under funding for FENA. Support from the Swiss National Science Foundation (SNSF) is also gratefully acknowledged.
\end{acknowledgments}

\appendix
\section{ Derivation of the winding exciton-photon coupling}

 Applying the minimal coupling substitution ${\bf k}\rightarrow{\bf k}+e\hat{{\bf A}}$ to the second-quantized representation of the quantum-well Hamiltonian (\ref{eq:quantum well}) leads to an electron-photon coupling of the form
\begin{equation}
	\hat{H}_{\mathrm{el-ph}}=
	-\mathrm{i}A\sum_{\mathbf{q},\mathbf{k}}F_{\mathbf{q}}\frac{1}{q}\left[q_{x}-\mathrm{i}q_{y}\right]
	\hat{c}_{c\mathbf{k+q}}^{\dagger}\hat{c}_{v\mathbf{k}}\hat{a}_{\mathbf{q}}
	+\mathrm{H.c.}
	\label{eq:H_el-ph}\,,
\end{equation}
where we have focused on the region around the exciton-photon resonance. Note that we have replaced the electron operators corresponding to the original angular-momentum basis (see Eq.~(\ref{eq:quantum well})) by conduction- and valence-band operators. Both basis are related by a band-mixing-induced rotation of the order of $Ak/M$ and are equivalent under the assumption that $A/M$ is small as compared to the Bohr radius $a_0 \sim 1/k$ of the excitons. We have verified numerically that the winding remains unchanged for larger $A/(Ma_0)\sim 1$ (for excitons with s-wave character). In that case, however, the amplitude of the exciton-photon coupling becomes suppressed.

To express the electron-photon coupling in terms of excitons, we invert
Eq.~\eqref{eq:exciton-1} as $c_{c,\mathbf{k}+\mathbf{q}/2}^{\dagger}c_{v,\mathbf{k}-\mathbf{q}/2}=\sum_{n}\phi_{n}^{*}(\mathbf{k})\hat{b}_{\mathbf{q}}^{(n)\dagger}$.
Focusing on energies close to the lowest-lying exciton allows for
dropping terms including higher exciton modes with $n>1$. With the
two-dimensional Fourier transform of the s-wave wavefunction $\phi_{1}(k)=2\sqrt{2\pi}a_{0}\left[1+(ka_{0})^{2}\right]^{-3/2}$,
we then obtain Eq.~\eqref{eq:winding_g} of the main text.

Note that the apparent non-analyticity of Eq.~\eqref{eq:winding_g} at
$q=0$ is an artifact of focusing only on the TE mode. For a
small in-plane momentum component, the TM-polarized modes can no longer
be neglected. In fact, the TE and TM polarization modes have equally
large in-plane electric fields if the momentum points predominately
in the $z$-direction. In this regime the basis of right- and left-handed circular polarizations is more suitable, and one finds a large constant coupling to the, say, right-handed mode, while the left-handed mode has a coupling with a double winding vanishing as $q^2$ (See Supplemental Material for more details. The momentum-depended switch from linear to circular polarization was also discussed in Ref.~\cite{solnyshkov_magnetic_2008}).  One can focus on the exciton-TE-photon coupling when the exciton is resonant with the TE mode at frequencies where the TM mode is gapped.
\\

\section{Topological gap opening for a two-dimensional Brillouin zone}
Assuming a square Brillouin zone, a gap of size $V_{{\rm X}}$ opens at the point $X=(\pi/a,0)$ similarly as in Fig.~\ref{fig:schematic periodic potential}. The main deviation from the one-dimensional case is the finite energy difference between the polaritons at the points $X$ and $M=(\pi/a,\pi/a)$. For a global gap, $V_{{\rm X}}$ should be larger than $\omega_{M}^{P-}-\omega_{X}^{P-}$. Note that $V_{{\rm X}}$ also increases the repulsion between exciton bands  backfolded into the first Brillouin zone from regions of the spectrum 
far away from resonance. These backfolded excitons are in an energy range $\omega_{0}^{X}\pm2V_{{\rm X}}$. There is therefore a tradeoff between the large exciton-photon coupling required to separate the
lowest polariton band from the backfolded excitons (large $g_{M}^{2}/(\omega_{M}^{{\rm C}}-\omega_{0}^{{\rm X}})$ as compared to $V_{{\rm X}}$) and the large $V_{{\rm X}}$ required to open a global gap. To satisfy these requirements in the best possible way, the Brillouin zone should be as circular as possible (e.g., hexagonal) to minimize the energy difference between the points located at its boundary. For an optimal choice of $a,$ $V_{{\rm X}}$, and Brillouin-zone geometry, we expect the size of the gap to be determined by the smallest of the two values $V_{{\rm X}}$ and $g_{M}^{2}/(\omega_{M}^{{\rm C}}-\omega_{0}^{{\rm X}})$.
\\

\bibliography{topolaritons}

\end{document}